\documentclass[a4paper,twoside]{article}

\usepackage{epsfig}
\usepackage{subfigure}
\usepackage{calc}
\usepackage{amssymb}
\usepackage{amstext}
\usepackage{amsmath}
\usepackage{amsthm}
\usepackage{multicol}
\usepackage{pslatex}
\usepackage{apalike}
\usepackage{SCITEPRESS}
\usepackage[small]{caption}

\usepackage[T1]{fontenc}
\usepackage{mathrsfs}
\usepackage{dsfont}
\usepackage{eurosym}
\usepackage{bm}
\usepackage{url}
\usepackage{stmaryrd}
\usepackage{pgf}
\DeclareGraphicsExtensions{.ps,.eps,.pdf}
\usepackage[pdftex]{hyperref}
\usepackage{balance}
\usepackage{multirow}
\usepackage[textwidth=2.0cm,textsize=scriptsize]{todonotes}

\usepackage{subfigure}
\usepackage{tikz}
\usepackage{float}
\usetikzlibrary{shapes.geometric}
\usetikzlibrary{arrows}

\begin{document}

\newcommand{\C}{\mathcal{C}}

\newcommand{\U}{\mathbf{U}}
\newcommand{\Snc}{\mathbf{S}}
\renewcommand{\R}{\mathbf{R}}

\newcommand{\Nat}{\mathbb{N}}
\newcommand{\Z}{\mathbb{Z}}
\newcommand{\Real}{\mathbb{R}}
\newcommand{\Q}{\mathbb{Q}}

\newcommand{\X}[1]{\mathbf{X}\left(#1\right)}
\newcommand{\Y}[1]{\mathbf{Y}\left(#1\right)}

\newcommand{\Zed}{\mathbf{Z}}
\newcommand{\Lng}{\mathscr {L}}
\newcommand{\iFF}{\Leftrightarrow}
\newcommand{\niFF}{\nLeftrightarrow}
\newcommand{\SNC}{{\mathcal S}}
\newcommand{\TRG}{{\mathcal T}}
\newcommand{\zot}{$\mathds{Z}$ot}
\newcommand{\G}[1]{\mathbf{G}\left(#1\right)}
\newcommand{\Hist}[1]{\mathbf{H}\left(#1\right)}
\renewcommand{\F}[1]{\mathbf{F}\left(#1\right)}
\newcommand{\triple}[3]{(#1, #2, #3)}
\newcommand{\pair}[2]{(#1, #2)}
\newcommand{\siff}{\Leftrightarrow}
\newcommand{\A}{\mathcal{A}}
\newcommand{\aX}{\mathrm{X}}
\newcommand{\aY}{\mathrm{Y}}
\newcommand{\x}{\mathbf{x}}
\newcommand{\eqdef}{\stackrel{\mbox{\begin{tiny}def\end{tiny}}}{=}} 
\newcommand{\iFFdef}{\stackrel{\mbox{\begin{tiny}def\end{tiny}}}{\iFF}}
\newcommand{\step}[1]{\xrightarrow{#1}}

\newcommand{\pspace}{\textsc{PSpace}}

\makeatletter
\def\Eqlfill@{\arrowfill@\Relbar\Relbar\Relbar}
\newcommand{\longmodels}[1][]{\,|\!\!\!\ext@arrow 0359\Eqlfill@{#1}}
\makeatother

\newcommand{\symodels}{\longmodels{\mbox{\it{\tiny sym}}}}

\newcommand{\intervaLii}[2]{[#1,#2]}
\newcommand{\intervaLie}[2]{[#1,#2)}
\newcommand{\intervaLee}[2]{(#1,#2)}

\newcommand{\interval}[2]{\langle #1,#2 \rangle}

\newcommand{\set}[1]{\{ #1 \}}

\newcommand{\tsys}[1]{\mathcal{S}(#1)}

\newcommand{\lapp}[1]{\lfloor #1 \rfloor}
\newcommand{\happ}[1]{\lceil #1 \rceil}

\newcommand{\first}[2]{(H_{#1}\vee L_{#1}) \wedge(\neg(H_{#1}\vee L_{#1}) \Snc (#2))}

\newcommand{\todoMR}[2]{\todo[color=red!50,#1]{\textbf{MR:}#2}}
\newcommand{\todoPL}[2]{\todo[color=yellow!50,#1]{\textbf{PL:}#2}}
\newcommand{\todoMB}[2]{\todo[color=green!50,#1]{\textbf{MB:}#2}}

\newcommand{\pname}[1]{\ensuremath{\textit{#1}}}
\newcommand{\on}{\pname{on}}
\newcommand{\off}{\pname{off}}
\newcommand{\lon}{\pname{l}}
\newcommand{\test}{\pname{test}}
\newcommand{\resetc}{\pname{rst-c}}
\newcommand{\turnoff}{\pname{turnoff}}


\title{An LTL Semantics of Business Workflows with Recovery}

\author{\authorname{Luca Ferrucci\sup{1,2}, Marcello M. Bersani\sup{2} and Manuel Mazzara\sup{3}}
\affiliation{\sup{1} ISTI-CNR, Italy}
\affiliation{\sup{2}Dipartimento di Elettronica Informazione e Bioingegneria, Politecnico di Milano, Italy}
\affiliation{\sup{3} Innopolis University, Russia and ETH Z{\"u}rich, Switzerland}
\email{luca.ferrucci@isti.cnr.it, marcellomaria.bersani@polimi.it, m.mazzara@innopolis.edu.ru}
}

\keywords{Business Workflow, Recovery Framework, Formal Methods, Temporal Logic, Semantics}

\abstract{
We describe a business workflow case study with abnormal behavior management (i.e. recovery) and demonstrate how temporal logics and model checking can provide a methodology to iteratively revise the design and obtain a correct-by construction system. To do so we define a formal semantics by giving a compilation of generic workflow patterns into LTL and we use the bound model checker \zot{} to prove specific properties and requirements validity. The working assumption is that such a lightweight approach would easily fit into processes that are already in place without the need for a radical change of procedures, tools and people's attitudes. The complexity of formalisms and invasiveness of methods have been demonstrated to be one of the major drawback and obstacle for deployment of formal engineering techniques into mundane projects.
}

\onecolumn 
\maketitle 
\normalsize 
\vfill

\section{Introduction}
\label{section-intro}

Nowadays Internet access is widespread and people use the internet for a wide range of activities,
among others to purchase goods and services. In Europe in 2012, 75\% of individuals aged 16 to 74 had used the internet in the previous 12 months, and nearly 60\% of these reported that they had shopped online (data collected by http://epp.eurostat.ec.europa.eu/). Presently Internet purchases represents the most important way of doing E-business while older systems are either been canceled or improved
in such a way that they are able to run over the Internet infrastructure. 

Together with the emerging of E-business and the exigency of exchanging business messages between 
trading partners, the concept of business integration arose. Business integration is becoming 
necessary to allow partners to communicate and exchange documents as catalogs, orders, reports
and invoices, overcoming architectural, applicative, and semantic differences, according to the
business processes implemented by each enterprise. In order to be effective, a business integration solution must also deal with (non-functional) requirements such as dependability, security, availability and compatibility.

In this work, our focus will be limited to the dependability aspect of business integration and the analysis of recovery solutions. In particular, we will give a formal semantics to business workflows
enriched with abnormal behavior and recovery and will address a methodology to define and verify specific causal properties defined by designers. Causal properties allows to specify order relationships between events and activities.


\subsubsection*{On methods and tools}

Logics and model-checking have been successfully used in the last decades for modeling and verification of various types of hardware and software systems and have a stronger credibility in the scientific community when compared with other formalisms. We here give an LTL-based semantics of workflow execution and use \zot{} \cite{PMP08} model checker for requirement verification. By applying model-checking on the case study presented in this paper, we demonstrate the feasibility of the verification approach and how temporal logics can work for both modeling and verification of (simple but realistic) business workflows inclusive of exception handling. This work has to be intended as complementary to what has been done in \cite{kn:LucMaz07} where a similar problem was approached in term of Process Algebra. A comparison of approaches is left as future work.


Differently from several others formalizations only offering languages without methods -- for a detailed discussion see \cite{Mazzara2009} and \cite{Mazzara2011} -- our approach, together with software tools, aims at offering a complete practical toolkit for software and systems engineers working in the field of workflow design. Following the line of \cite{Mazzara2013} this approach goes under the correct-by-construction paradigm and the idea of developing dependable systems by integrating specific approaches well-suited to each development phase.

The ideal process of workflow verification is an iterative process. In this work, we aim at providing an instrument to designers for workflow revision, i.e. a procedure to follow until the requirements are finally met. To do this, we encode the workflow into a formal language and, at the same time, we formally describe specific requirements on the system. This is discussed in Section \ref{section-ltl}. At this point, as shown in Section \ref{section-experimental}, correctness can be automatically determined via the \zot{} model checker. As an outcome of model checking we may need to revise the workflow in order to meet the requirements.


\subsubsection*{A descriptive semantics}



The role of temporal logics in verification and validation is two-fold. First, temporal logic allows abstract, concise and convenient expression of required properties of a system. In fact, Linear Temporal Logic (LTL) is often used with this goal in the verification of finite-state models, e.g., in model checking~\cite{BK08}. Second, temporal logic can be used as a descriptive approach for specifying and modeling systems (see, e.g.,~\cite{MS94,FMMR12}). A descriptive model is based on axioms, written in some (temporal) logic, which define the system through its general properties, rather than by an operational model based on some kind of machine behaving in the desired way. In this case, verification typically consists of satisfiability checking of the conjunction of the model and of (the negation of) its desired properties. For example, in Bounded Satisfiability Checking (BSC) \cite{PMS12}, Metric Temporal Logic (MTL) specifications on {\em discrete} time and properties are translated into propositional logic, in an approach similar to Bounded Model Checking of LTL properties of finite-state machines.

Specifying temporal relations among events that do not inherently behave in an operational way, by using an operational model, may become rather hard. This is the case for the system recovery considered here. Exception handling is an event-based paradigm that implements the asynchronous exchange of warning events among actors that are part of the system. The typical implementation of exception handling mechanisms -- through logical rules of the form \emph{if $(cond)$ then \emph{throw}$(e)$} and \emph{try-catch} blocks-- requires ad-hoc extensions of operational-based formalisms by means of the definition of message-passing primitives.
Specifying exception handling mechanisms through temporal logic does not require extending LTL and also allows modeling of two sorts of exceptions (punctual and non-punctual, see Section~\ref{section-ltl}) in a coherent and uniform way by providing events that represent exceptions a suitable semantics.

\subsubsection*{Other approaches and novelty}

Several approaches have been adopted in recent years to provide formal semantics of business 
processes. Most of them are very much bound to a specific formalism accordingly extended to better cope with modeling issues. These attempts mostly belong (but not limited) to the process algebras, Petri nets or model-based philosophies, with some raid into temporal logics et similia too. 

Mobile process algebra have been successfully used in \cite{kn:LucMaz07} that this work intend to complement. Limitations of process algebras approaches like the previous ones and, for example, \cite{VazF12} are related to the fact that process algebras are based on equational reasoning. From a practical perspective, this makes verification tricky, difficult and certainly not user-friendly, because verification is mainly carried out by specific proof techniques that are used to prove behavioural equivalence among processes. Furthermore, all these approaches mostly focus on the verification of reachability-based properties (with some exceptions like \cite{CalzolaiNLT08}) and tool support is very limited (see \cite{Mazzara2010}). On the other side, other works like \cite{MontesiGZ14} provide a methodology and tool support for the modeling phase, but do not cope with the verification phase and either do not belong to the correct-by-construction paradigm.

Petri Nets supporters and van der Aalst approaches like Workflow Petri Nets (WPN) \cite{Aalst:1997} reached the objective of verification and tool support to a much larger extent than other communities. This approach is based on extensions of previously existing formalisms and still represents an operational model, which also inherits the relative overhead. A successful attempt to overcome this issue has been provided by \cite{YYT09} where acyclic WPN are translated into a finite-state automaton and verified against a suitable LTL property in order to verify soundness.

Model-based approaches have also been used, though to a much lesser extent and often in combination with testing, for validation of business critical systems. The B-model is one of the most popular together with its reactive-systems extension Event-B \cite{AugustoHGFGL03}. B and Event-B are not lightweight methods. They do come with a refinement-based methodology, but cannot easily be embedded into already existing industrial processes \cite{Mazzara2013}. 

In the domain of temporal logics, CTL has been used to specify and enforce intertask dependencies \cite{Attie93specifyingand}, and LTL for UML activity graphs verification \cite{Eshuis02verificationsupport}.
Other temporal logics have also been used for similar objectives. In paricular, in \cite{BMMR12} a complete and coherent semantics based on the TRIO logic \cite{Ghezzi:1990} has been proposed for a more consistent set of UML diagrams.


Recovery frameworks have been more rarely formalized in similar manners instead. This has to be intended as another major contribution of the paper. In \cite{Eisentraut:2009} the state-of-the-art 
in formalizing fault, compensation and termination mechanisms of WS-BPEL 2.0 has been deeply investigated. One of the first works formally discussing business recovery in terms of long-running transactions is \cite{ButlerF04}.

The working assumption is that a lightweight solution would easily fit into processes that are already in place without the need for a radical change of procedures, tools and people's attitudes, which is actually the case for most of the aforementioned techniques. The complexity of formalisms and invasiveness of methods have been demonstrated to be one of the major drawback and obstacle for deployment of formal engineering techniques into mundane projects \cite{Mazzara2013}, \cite{RomanovskyThomas2013}.


The rest of the paper is organized as follows: Section \ref{section-casestudy} describes the case study of a workflow for order processing. The semantics of workflows and exception handling is given using temporal logic in Section \ref{section-ltl} where a general encoding into LTL is provided. In Section \ref{section-experimental} the implementation of this translation is illustrated and tests have been carried out to validate its correctness. Finally, Section \ref{section-conclusions} draws conclusive remarks and focus on future developments.

\section{Workflows with Recovery}
\label{section-casestudy}

A business process is a set of logically related tasks performed to achieve a well defined
business outcome. Examples of typical business processes are elaborating a credit claim, 
hiring a new employee, ordering goods from a supplier, creating a marketing plan, processing 
and paying an insurance claim, and so on. Many computer systems are already available in the 
commercial marketplace  to address the various aspects of Business Process Management (BPM) 
and automation.

An automated business process is generally called \textit{business workflow}, i.e a choreographed 
and system-driven sequence of activities directed towards performing a certain business task to completion. For activity we intend an element that performs a specific function within a process. Activities can be as simple as sending or receiving a message, or as complex as coordinating the 
execution of other processes and activities. A business process may encompass complex activities 
some of which run on back-end systems such as, for example, a credit check, automated billing, a purchase order, stock updates and shipping, or even such frivolous activities as sending a document
and filling a form.

Workflow is commonly used to define the dynamic behavior of business systems and originates from business and management as a way of modeling business processes that could wholly or partially be automated. It has evolved from the notion of process in manufacturing and offices because these processes are the result of trying to increase efficiency in routine work activities since industrialization.

The view on a workflow inherited from the BPM perspective -- i.e. the way in which workflow designers may see a system -- is somehow different from the way formalists see it. Therefore, to fill the gap between the formal and informal world, we will provide the reader with a precise understanding.

A workflow is a direct graph defined by pair $(P, E)$, where $P$ is a finite non empty set of places and $E$ is a relation, called flow of execution, that is defined as $E \subseteq P \times P$.
Elements of $E$ are pairs $(p,q)$, with $p,q\in P$, that are called \emph{transitions}.
Let $A$ be a place of $P$.
Set $\mathit{out}(A)$ is the set of \emph{outgoing transitions} starting from $A$ which is defined as $\{(A,q) \mid q\in P, (A,q) \in E\}$.
Set $\mathit{in}(A)$ is the set of \emph{ingoing transitions} leading to $A$ which is defined as $\{(q,A) \mid q \in P, (q,A)\in E\}$. 

We assume that $|\mathit{out}(A)|\geq 1$, for all $A \in P$, except for place $\mathit{end}$, and that $|\mathit{in}(A)|\geq 1$, for all activity $A$, except for place $\mathit{start}$. 

A \emph{finite path} from $p$ to $p'$ is a (finite) sequence of pairs $(p_0,p_1) \dots (p_{n-1},p_n)$ with $p_0=p$ and $p_n=p'$, such that $(p_{i},p_{i-1}) \in E$, for all $0\leq i\leq n$. 
An \emph{infinite path} from $p$ is an (infinite) sequence of pairs $(p_{i},p_{i-1}) \in E$, for all $i\geq 0$, where $p_0=p$.
Throughout the paper, we assume that workflows are structurally correct, that is, such that there exists at least one path from place $\mathit{start}$ to (any) place $\mathit{end}$.
Informally, an \emph{execution} of a workflow is the superposition of paths of the workflow starting from the initial place.

Each activity may have a specific semantics that forces the execution flow to be compliant with some specific rules.
However, endowing activities with a semantics is not achievable, in general, only by considering workflows as graphs and the definition of complex behaviors may require more specific and expressive formalisms.
This is the case of \textit{conditional cases} and \textit{split-join} activities that we consider in this paper whose semantics can be easily obtained by defining concise LTL formulae.
Conditional cases model \textit{if-then-else} blocks provided with the usual semantics.
If the condition holds the ``then'' branch is executed otherwise the execution flow follows the ``else'' branch.
In this paper, we do not model guards explicitely, though conditional expressions over finite domains can be easily introduced, as the effort of the work is focused on the exception recovery mechanism.
Split-join activities model the parallel execution of two (or more) branches of the workflow that starts concurrently when activity \textit{split} is executed and eventually synchronize their computations in correspondence to the associated \textit{join} activity.
We assume that conditional cases and split-join are fictitious activities with non relevant time duration.

Workflows are endowed with \emph{exceptions}, i.e., specific events (or signals) representing erroneous configurations that occur during the execution and that may prevent the workflow from reaching a final place.
With no loss of generality, we assume that an exception (raised at some moment throughout the execution) that is not managed by the workflow, forces the running activities that monitor the exception not to terminate.
Under such assumption, the termination of an execution, and then of all the activities occurring therein, can only be guaranteed if $\mathit{end}$ is reached.
However, the assumption does not prevent modeling an activity, say $A$, that terminates with an error configuration.
In fact, one can introduce an exception to represent the wrong termination of $A$ and a special activity that is able to detect it and specifically devised for managing faulty termination of $A$.
In addition, workflow executions are not restricted only to finite paths (from $\mathit{start}$ to $\mathit{end}$) and infinite iterations of  finite paths of the workflow are still allowed.
In fact, infinite executions are representative of wrong behaviours only when there is one (ore more) activity, over some paths, that can not terminate and does not allow the workflow to proceed further and reach $\mathit{end}$.
To guarantee that a workflow is correctly designed, all the exceptions that may raise during an execution have to be caught and solved.
Designers should prevent such anomalous situations by defining suitable recovery actions to be executed simultaneously with the activity running on error and that restore the workflow execution.

\newcommand{\catch}{\mathit{catch}}
\newcommand{\probe}{\mathit{probe}}
\newcommand{\throw}{\mathit{throw}}

Let $E$ be a (finite) set of exceptions associated with the workflow and $P$ and $S$ be two subsets of $E$ such that $P\cup S = E$ and $P \cap S = \emptyset$.
Set $P$ is the set of permanent (i.e., non-punctual) exceptions and $S$ is the set of punctual exceptions.
Activity $start$ and $end$ are not associated with any exception.
Informally, we say that an exception is punctual when
its duration is negligible.
Conversely, an exception is non-punctual when it may have a duration and it lasts from a position where it is raised until a position where it expires.
Each activity $A \in W$ can be associated with three, possibly empty, sets of exceptions.
Set $\throw(A)$ is the set of exceptions that activity $A$ can notify whenever a potential dangerous error may compromise the workflow execution and that have to be suitably handled by some other activity which is able to repair the fault.
Set $\catch(A)$ contains exceptions that activity $A$ can handle, that is, that the activity may take on responsibility of remedying the fault.
Set $\probe(A)$ is the set of exceptions that may compromise the workflow execution because they let activity $A$ switch to an error state, if no activity catching them is active at the same time.

In Section \ref{section-ltl}, we define formally, through LTL formulae, the semantics of correct workflows executions with respect to the specific semantics of conditional cases and split-join activities and recovery exceptions.

\begin{figure}
\centering
\includegraphics[scale=0.4]{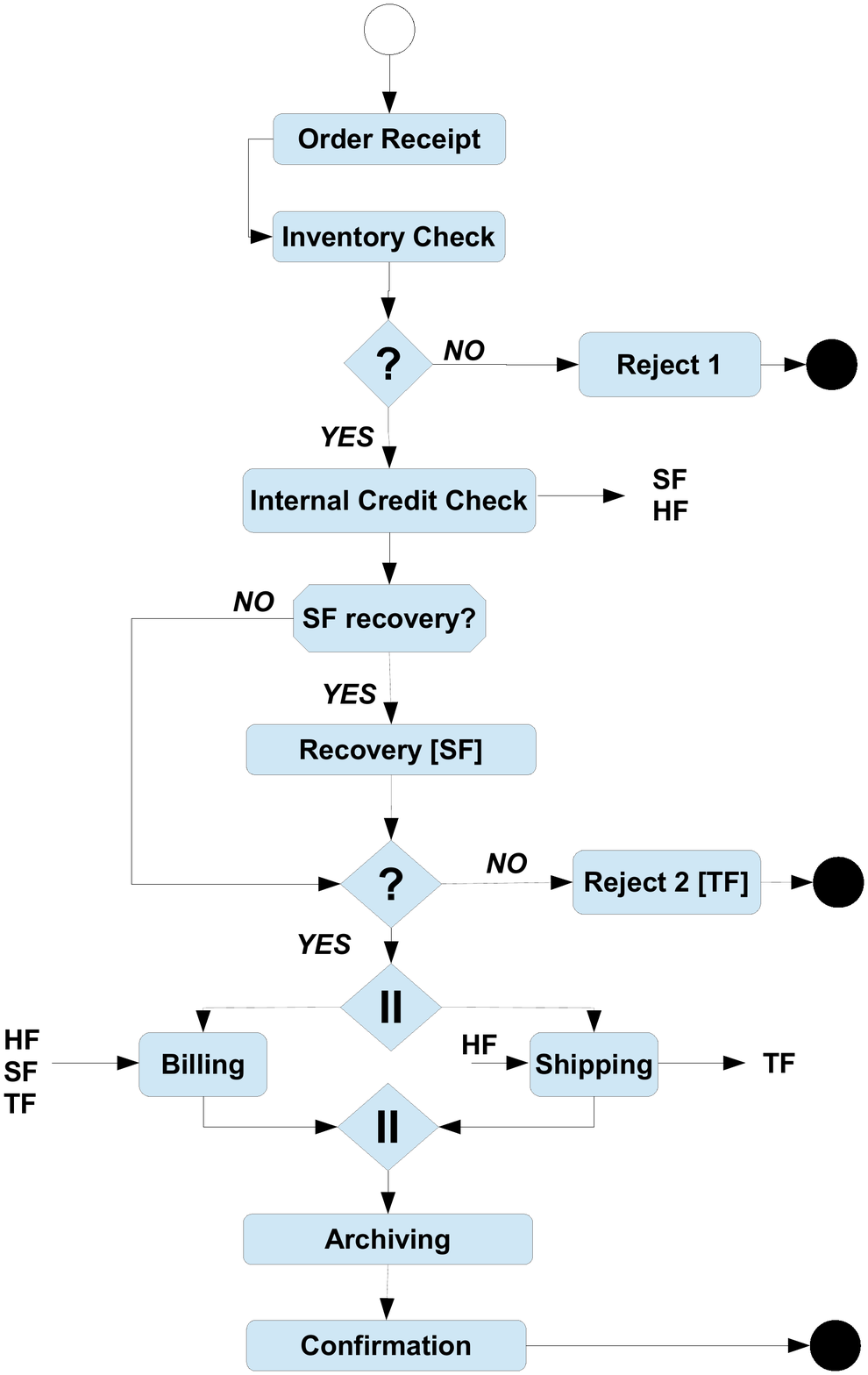}
\caption{Workflow Case Study}
\label{fig:workflow}
\end{figure}

The case study approached in this paper is depicted in Figure~\ref{fig:workflow} and describes a typical office workflow for order processing that is commonly found in large and medium-sized 
organizations (for more details \cite{Ellis:1995}). Although the example 
may appear too simple, most of the online purchase systems are actually of 
comparable complexity, apart abstracting from several details. 
The workflow we provide in Figure~\ref{fig:workflow} is a simple example of wrong design as some exceptions are handled incorrectly and may cause infinite executions.
To demonstrate the effectiveness of our approach, in Section~\ref{section-experimental}, we verify (the LTL model of) the workflow and we show that discovering wrong executions allows us to enforce model refinement and obtain a correct design.
The workflow consists of the ten activities, depicted within rectangles, and is endowed with two permanent exceptions \textit{HF} (Hardware Failure), \textit{SF} (Software Failure) and one punctual exception \textit{TF} (Transport Failure).
An ingoing arrow into an activity labelled with exceptions defines the set $\probe$ of the activity while an outgoing arrow, labelled with exceptions, defines the set $\throw$.
Concerning the workflow in Figure~\ref{fig:workflow}, we have $\throw(\mathit{Internal Credit Check})=\{\mathit{HF},\mathit{SF}\}$ and $\throw(\mathit{Shipping})=\{\mathit{TF}\}$.
The set of exceptions that an activity catches is defined within square brackets and it is written beside the name of the activity in the rectangle.
We have $\catch(\mathit{Recovery})=\{\mathit{SF}\}$ and $\catch(\mathit{Reject_2})=\{\mathit{TF}\}$.
Conditional cases are graphically represented by diamonds, and labelled with $?$ and \textit{SF recovery} in the workflow of Figure~\ref{fig:workflow}. 
Split-join activities are depicted by diamonds labelled with $\rVert$. 


\section{Formal Semantics}\label{section-ltl}



LTL~\cite{LPZ85} is one of the most popular descriptive language for defining temporal behaviors that are represented as sequences of observations.  The time model adopted in this logic is a totally ordered set (e.g., $(\mathbb{N},<)$) whose elements are the positions where the behavior is observed. 
LTL allows the expression of positional orders of events both towards the past and the future.
Let $AP$ be a finite set of atomic propositions. Well-formed LTL formulae are defined as follows:
\begin{equation*}
  \phi :=
  \begin{gathered}
    p \mid \phi \wedge \phi \mid \neg \phi \mid   \X{\phi} \mid \Y{\phi} \mid \phi\U\phi \mid \phi\Snc\phi
  \end{gathered}
\end{equation*}
where $p \in AP$, $\mathbf{X}$, $\mathbf{Y}$, $\U$ and $\Snc$ are the usual ``next'', ``previous'', ``until'' and ``since'' modalities.
The dual operator ``release'' $\R$  is defined as usual, i.e., $\phi\R \psi$ is $\neg(\neg \phi \U \neg\psi)$.
Useful operators can be defined from the previous ones. 
``Eventually'' is defined as $\F{\phi}=\mathit{true}\U\phi$ and ``Globally" $\mathbf{G}\phi$ operator is  $\mathit{false} \R \phi$. 
Informally, $\F{\phi}$ means that $\phi$ will eventually occur in the future, including the current position, and $\G{\phi}$ means that $\phi$ holds indefinitely from the current position.

The semantics of LTL formulae is defined with respect to a strict linear order representing time $\pair{\Nat}{<}$. Truth values of propositions in $AP$ are defined by interpretation $\pi :\Nat \to \wp(AP)$ associating a subset of the set of propositions with each element of $\Nat$. The semantics of a LTL formula $\phi$ at instant $i\geq 0$ over a linear structure $\pi$ is recursively defined as in Table~\ref{tab:LTLsemantics},
\begin{table}[h!]
\begin{equation*}
\small
\begin{aligned}
\pi, i &\models p  \iFF  p \in \pi(i) \text{ for } p \in AP \\
\pi, i &\models \X \phi \iFF \pi,i+1 \models \phi \\
\pi, i &\models \Y \phi \iFF \pi,i-1 \models
\phi \wedge i>0\\
\pi, i &\models \phi\U\psi \iFF
\exists \, j\geq i: \pi,j \models \psi \ \wedge \pi,n \models \phi \ \forall \ i\leq n < j \\
\pi, i &\models \phi\Snc\psi \iFF
\exists \, 0\leq j \leq i: \pi,j \models \psi \, \wedge \pi,n \models \phi \ \forall \ j < n \leq i \\ 
\end{aligned}
\end{equation*}
\caption{Semantics of LTL. Boolean connectives are omitted for brevity.}\label{tab:LTLsemantics}
\end{table}
A formula $\phi \in$ LTL is \emph{satisfiable} if there exists an interpretation $\pi$ such that $\pi,0 \models \phi$. 

\begin{table*}[ht]
\small
\centering
\begin{tabular}{c|c|c}
$[A]_{t_{out}(A)}$ &
\begin{minipage}{7cm}
\begin{gather}
A \Rightarrow (A \land \neg t_{\mathrm{out}}(A)) \U (t_{\mathrm{out}}(A))\label{f-act-followed-trans-1} \lor \G{A}\\
\bigwedge_{i\in \mathit{out}(A)} (t_i \Rightarrow \Y{A} \land \neg A)\label{f-act-followed-trans-2} %
\end{gather}
\end{minipage} & 
\begin{minipage}{2cm}
\begin{tikzpicture}
\draw (0,0) node[rectangle,draw] {$A$};
\draw[->]  (0.47,0) -- (1,0);
\draw[->]  (-0.47,0) -- (-1,0);
\draw[->]  (0,-0.5) -- (0,-1);
\draw  (0.8,0.2) node {$t_2$};
\draw (-0.8,0.2) node {$t_1$};
\draw (0.2,-0.8) node {$t_i$};
\end{tikzpicture}
\end{minipage}\\
\hline 
${_{t_{in}(A)}}[A]$ & 
\begin{minipage}{7cm}
\begin{gather}
A \Rightarrow (A \land \neg t_{\mathrm{in}}(A)) \Snc (t_{\mathrm{in}}(A))\label{f-trans-followed-act-1} \\
\bigwedge_{i\in \mathit{in}(A)} (t_i \Rightarrow \X{A} \land \neg A)\label{f-trans-followed-act-2}
\end{gather}
\end{minipage} & 
\begin{minipage}{2cm}
\begin{tikzpicture}
\draw (0,0) node[rectangle,draw] {$A$};
\draw[<-]  (0.47,0) -- (1,0);
\draw[<-]  (-0.47,0) -- (-1,0);
\draw[<-]  (0,0.5) -- (0,1);
\draw  (0.8,0.2) node {$t_2$};
\draw (-0.8,0.2) node {$t_1$};
\draw (0.2,0.8) node {$t_i$};
\end{tikzpicture}
\end{minipage}\\
\hline
$[\cdot+\cdot]$ & 
\begin{minipage}{7cm}
\begin{gather}
t_1 \Rightarrow \neg t_2 \label{f-cond-1}\\
\oplus \Rightarrow \neg \Y{\oplus} \land \neg \X{\oplus}\label{f-cond-2}
\end{gather}
\end{minipage} & 
\begin{minipage}{2cm}
\begin{tikzpicture}
\draw (0,0) node[diamond,draw] {+};
\draw[->]  (0.4,0) -- (1,0);
\draw[->] (-0.4,0) -- (-1,0);
\draw (0.8,0.2) node {$t_1$};
\draw (-0.8,0.2) node {$t_2$};
\end{tikzpicture}
\end{minipage}\\
\hline
$[\cdot\mid \cdot]$ & 
\begin{minipage}{7cm}
\begin{gather}
\bigwedge_{t_1,t_2 \in \mathit{out}(\rVert)} (t_1 \iFF  t_2) \text{ or }\bigwedge_{t_1,t_2 \in \mathit{in}(\rVert)} (t_1 \iFF  t_2)\label{f-split-1}\\
\rVert \Rightarrow \neg \Y{\rVert} \land \neg \X{\rVert}\label{f-split-2}
\end{gather}
\end{minipage} & 
\begin{minipage}{5cm}
\vspace{10pt}
\begin{tabular}{cc}
\begin{tikzpicture}
\draw (0,0) node[diamond,draw] {$\rVert$};
\draw[->]  (0.47,0) -- (1,0);
\draw[->]  (-0.47,0) -- (-1,0);
\draw[->]  (0,-0.5) -- (0,-1);
\draw  (0.8,0.2) node {$t_2$};
\draw (-0.8,0.2) node {$t_1$};
\draw (0.2,-0.8) node {$t_i$};
\end{tikzpicture}
&
\begin{tikzpicture}
\draw (0,0) node[diamond,draw] {$\rVert$};
\draw[<-]  (0.47,0) -- (1,0);
\draw[<-]  (-0.47,0) -- (-1,0);
\draw[<-]  (0,0.5) -- (0,1);
\draw  (0.8,0.2) node {$t_2$};
\draw (-0.8,0.2) node {$t_1$};
\draw (0.2,0.8) node {$t_i$};
\end{tikzpicture}
\end{tabular}
\end{minipage}
 \\
\end{tabular} 
\caption{Workflow LTL encoding}
\label{tab:translation}
\end{table*}

\subsubsection*{Workflow model}

Workflows model execution of systems as sequences of activities. Transitions, conditional cases and split-join interleave the activities and determine uniquely the flow of the execution, i.e., the sequence of activities that realizes the computation. An activity is an abstraction of a compound of actions that are performed by the real workflow. Although they can be modelled as atomic computations, we adopt a different perspective for which the activities, being actions in the real world, have a non-punctual duration. To translate workflows into an LTL formula, we assume that an activity is always followed by a transition, and viceversa, and that conditional tests and splits are special activities that have punctual duration. When an activity is performed, the firing of the outgoing transition lets the system change allowing it to execute the next activity. Therefore, our LTL translation allows modelling of workflows as sequences of activities and transitions, in strict alternation.

With no loss of generality, we assume that no element in the same graph is duplicated. By this assumption, we associate each element with an atomic proposition that uniquely identifies it.
We write $A$ to represent the activity $A$ in the workflow, while we write $t$ to indicate a transition between two activities. If $A$ holds at position $i$ then the workflow is performing activity $A$ at that position; similarly for $t$. We introduce $\rVert$ and $\oplus$ to indicate the split and the conditional activity, respectively; $\mathit{start}$ and $\mathit{end}$ to indicate the starting and the final activity of the workflow. Workflow diagrams are translated according to rules in Table~\ref{tab:translation}.


Let $t_{\mathrm{out}}(A)$ be the disjunction $\bigvee_{t_i \in \mathit{out}(A)} t_i$ and $t_{\mathrm{in}}(A)$ be the disjunction $\bigvee_{t_i \in \mathit{in}(A)} t_i$.

Formula~\eqref{f-act-followed-trans-1} states that if activity $A$ holds at the current position, then it is true, at that position, that either $A$ holds forever, which is the case of a workflow that is blocked because of an exception, or $A$ but not $t_{\mathrm{out}}(A)$ holds until at least one transition in $\mathit{out}(A)$ holds. In this way, activity $A$ lasts until one of its outgoing transition fires. Formula~\eqref{f-act-followed-trans-2} imposes that if transition $t_i \in \mathit{out}(A)$ holds at position $i$, then in the previous one $i-1$, activity $A$ holds. This necessary condition enforces that a transition fires only if the activity from which it originates has just been terminated.

Formula~\eqref{f-trans-followed-act-1} states that if activity $A$ holds at the current position, then it is true, at that position, that $A$ but not $t_{\mathrm{in}}(A)$ holds since at least one transition in $\mathit{in}(A)$ has been fired. In this way, activity $A$ has lasted since one of its ingoing transition fired. Formula~\eqref{f-trans-followed-act-2} imposes that if transition $t_i \in \mathit{in}(A)$ holds at position $i$, then, in the next one $i+1$, activity $A$ holds. This necessary condition enforces that a transition fires only if the activity to which it leads will be performed in the next position.

The translation for $[A]_{t_{out}(A)}$ and ${_{t_{in}(A)}}[A]$ does not impose constraints on transitions in $t_{out}(A)$ and $t_{in}(A)$. In the case of $[A]_{t_{out}(A)}$, the execution flow after an activity $A$ is nondeterministically defined as it is determined by at least one transition that fires when $A$ terminates. Simmetrically, the execution of an activity $A$ that is reached by more than one transition in ${_{t_{in}(A)}}[A]$ can be started by at least one execution that is active before $A$. The conditional case $[\cdot + \cdot]$ and the split-join $[\cdot\mid\cdot]$ activities are modelled by using the rules (1)-(4) and, in addition, specific constraints to enforce the proper flow of execution.
To model the flow of the conditional cases, we force the execution of the two branches to be exclusive as for the \textit{if-then-else} construct of programming languages. The translation of the split (resp. join) activity is similar yet it enforces the synchronization of all the transitions starting from (resp. yielding to) it.

The conditional case $[\cdot\mid\cdot]$ is translated compositionally. For each conditional activity we introduce a new fresh atomic proposition $\oplus$. Formulae \eqref{f-act-followed-trans-1}, \eqref{f-act-followed-trans-2}, \eqref{f-trans-followed-act-1} and \eqref{f-trans-followed-act-2} rule the semantics the same as sequence of actions. In addition, Formula \eqref{f-cond-1} impose that only one branch is executed, by forcing the strict complementarity between $t_1$ and $t_2$. Formula \eqref{f-cond-2} enforces punctuality of $\oplus$ and states that if $\oplus$ holds at position $i$ then in the next and in the previous positions it does not hold.

The translation for the split-join activities is similar to the one defined for the conditional case.
For each split activity we introduce a new fresh atomic proposition $\rVert$ and we use formulae \eqref{f-act-followed-trans-1}, \eqref{f-act-followed-trans-2}, \eqref{f-trans-followed-act-1} and \eqref{f-trans-followed-act-2} to rule the semantics as an activity plus Formula \eqref{f-cond-2} which enforces punctuality of $\rVert$, similarly to the previous case. The only difference is Formula \eqref{f-split-1} that is divided into two parts that are used exclusively. Both of them impose strict contemporaneity of all the transitions involved in activity $\rVert$. The first one is defined only for the split activity and states that all the outgoing transitions starting from it, occur at the same time.
The second one is similar but for the join activity, where all the parallel computations must join before proceeding further. It states that all the ingoing transitions leading to it, occur at the same time.

\newcommand{\ship}{\mathit{Ship}}
\newcommand{\bill}{\mathit{Bill}}
\newcommand{\arch}{\mathit{Arch}}
\newcommand{\conf}{\mathit{Conf}}
\newcommand{\rej}{\mathit{Rej}}
\newcommand{\eendwf}{\mathit{end}} 
We assume that when a workflow terminates, it never resumes, by adding to the model formula $\eendwf \Rightarrow \G{\eendwf}$.
The assumption is realistic because business process executions are unique and always have a starting point, where inputs are collected and fed to the process, and terminate by producing an outcome.
Infinite repetitions of finite (correct) executions of a workflow are not meaningful for our purpose as our intention aims at modeling infinite executions only when they represent wrong exception handling.

\subsubsection*{Encoding exceptions}
Let $E$ be a (finite) set of exceptions associated with the workflow and $P$ and $S$ be two subsets of $E$ such that $P\cup S = E$ and $P \cap S = \emptyset$ where $P$ is the set of permanent exceptions and $S$ is the set of punctual exceptions.
Observe that $start$ and $end$ are not associated with any exception and then, they are not considered as activity.

Informally, we say that an exception is punctual when it holds exactly one time instant whenever it occurs.
Conversely, an exception is non-punctual when it may have a duration and it lasts from a position where it is raised until a position where it expires.
Let $s \in S$. 
To model punctual exception $s$ we introduce the following Formula \ref{f-punctual-ex} that forces exception $s$ to be false in the next position of the one where exception $s$ occurs.
\begin{equation}\label{f-punctual-ex}
\bigwedge_{e \in S} (e \Rightarrow \neg \X{e})
\end{equation}

Let $A$ be an activity and $\catch(A)$ be the set of exceptions that activity $A$ can restores.
Non-punctual exceptions may hold continuously over some adjacent positions.
When such an exception occurs, at some position, then it holds until an activity $A$ such that $e \in \catch(A)$ restores the exception.
The following Formula~\ref{f-non-punctual-ex} states that if, at the current position, $e$ holds then there is a position in the future where an activity restores it, otherwise it will hold indefinitely.
In fact, the consequent of the implication imposes that if $\bigvee_{\stackrel{A}{e \in \catch(A)}}(e \U A)$ holds then $\neg \G{e}$ must hold, that is, $e$ will not hold indefinitely.
Conversely, if $\bigvee_{\stackrel{A}{e \in \catch(A)}}(e \U A)$ does not hold then $\neg \G{e}$ must not holds, that is, $e$ will hold indefinitely.
\begin{equation}\label{f-non-punctual-ex}
\bigwedge_{e \in P} (e \Rightarrow 
(\neg \G{e} \iFF \bigvee_{\stackrel{A}{e \in \catch(A)}}(e \U A))
\end{equation}

Let $A$ be an activity and $\probe(A)$ be the set of exceptions associated with $A$ that may let $A$ loop indefinitely.
If $A$ is active at a certain position of the time, then the occurrence of an exception $e$ in $\probe{}(A)$ causes an abortion of $A$ if, at that moment, there is no activity $B$ that restores $e$, such that $e\in \catch(B)$.
The abortion represents a configuration of error that can not be restored, i.e., $A$ loops infinitely or terminates with a system error.
Formula \ref{f-activity-error} states that, if at the current position, activity $A$ holds and exception $e$ occurs and no activity managing $e$ is active, i.e., $e \in \catch(B)$, then activity $A$ will never terminate.
\begin{equation}\label{f-activity-error}
\bigwedge_{e\in\probe(A)} (A\land e \land \bigwedge_{\stackrel{B}{e \in \catch(B)}}\neg B) \Rightarrow \G{A}
\end{equation}
Formula \ref{f-activity-error} is defined for all activities $A$ of the workflow with a non empty set $\probe(A)$.

Following Formulae~\ref{f-activity-error-nec} and \ref{f-activity-error-nec-II} define the necessary conditions to have infinite execution.
Formula~\ref{f-activity-error-nec} is specific for punctual exceptions.
At a certain position, if activity $A$ is active and it never terminates, i.e., $\G{A}$ holds at that position, then there exists an activity $C$ of the workflow, possibly different from $A$, that eventually loops indefinitely because an exception $e\in \probe(C)$ is not correctly handled.
This allows modelling the fact that an infinite execution of an activity may be enforced by a different activity that goes into an error state.
Moreover, the faulty activity $C$ may start its execution when activity $A$ is already running and the occurrence of the exception that induces the infinite looping error state of $C$ may occur even later its starting position.
This explains the $\mathbf{F}$ in the consequent of the formula that holds when there is a position $i$, possibly following the position where $\G{A}$ begins to hold,
such that, from that position $i$, there is a position in the past throughout the execution of $C$ where an exception $e\in \probe(C)$ occurred and no activity managing $e$ was active.
Let $W$ be the set of activity defining the workflow.
\begin{equation}\label{f-activity-error-nec}
\G{A} \Rightarrow  \F{\bigvee_{C\in W} (C) \Snc(C\land \bigvee_{e\in \probe(C)} (e \land \bigwedge_{\stackrel{B}{e \in \catch(B)}}\neg B))}
\end{equation}
Formula~\ref{f-activity-error-nec} does not apply to non-punctual exception because $\Snc$ may hold only in one position (an this is enough to have $\G{A}$) because non-punctual exceptions are not forced to hold indefinitely when no activity of the workflow can eventually handle them.
Next formula~\ref{f-activity-error-nec-II} remedies the problem and requires that if activity $A$ holds forever, then there is an activity $C$ (which may possibly be $A$) and a non-punctual exception $e\in \probe(C)$ that holds indefinitely, because no activity $B$ ever catches $e$.
\begin{equation}\label{f-activity-error-nec-II}
\G{A} \Rightarrow  \F{\bigvee_{\stackrel{C\in W}{e\in \probe(C)}} \G{C \land e \land \bigwedge_{\stackrel{B}{e \in \catch(B)}}\neg B)}}
\end{equation}
Both formulae \ref{f-activity-error-nec} and \ref{f-activity-error-nec-II} are defined for all activities $A$ appearing in the workflow.
Observe that when $\probe(C)$, for some $C\in W$, is empty then the second formula of $\Snc$, in Formula~\ref{f-activity-error-nec}, and the formula within $\mathbf{F}$, in Formula~\ref{f-activity-error-nec-II}, are trivially false.
In this case, the activity appearing in the antecedent of the formula always terminates and no looping executions are admitted for it, because $\G{A}$ is false.

An exception $e\in E$ is \emph{internal} if it is thrown by some activity appearing in the workflow whereas it is \emph{external} otherwise.
Next Formula~\ref{f-throw-iexc-nec} defines the necessary condition so that an internal exception is thrown. 
Let $\throw (A)$ be the set of exceptions that activity $A$ may rise.
The first formula states that if exception $e\in S$ holds then there exists an activity $A$ that is active at the same position such that $e$ belong to the set of exceptions which $A$ can raise, i.e., $e\in\throw(A)$.
The second formula requires that if a permanent exception $e$ holds then there is an activity $B$ and a position in the past where $B$ raised $e$.
\begin{equation}\label{f-throw-iexc-nec}
\begin{gathered}
\bigwedge_{e \in S}(e \Rightarrow \bigvee_{\stackrel{A}{e\in\throw(A)}} A)\\
\bigwedge_{e \in P}(e \Rightarrow (e) \Snc (e \land \bigvee_{\stackrel{B}{e \in \throw(B)}}B)).
\end{gathered}
\end{equation}
Next Formula~\ref{f-throw-eexc-nec} defines the necessary condition for external exceptions. 
It only requires that when an external exception occurs then there is an  activity that is active at some position in the past including the current one; i.e., they can not happen only in correspondence of transitions.
\begin{equation}\label{f-throw-eexc-nec}
\begin{gathered}
\bigwedge_{e \in S}(e \Rightarrow \bigvee_{A \in W} A)\\
\bigwedge_{e \in P}(e \Rightarrow (e) \Snc (e\land \bigvee_{B\in W}B)).
\end{gathered}
\end{equation}

\section{Experimental Results}
\label{section-experimental}

In this section, we provide experimental evidences regarding the proposed approach. The evaluation has been conducted by executing and elaborating a series of tests, which have been carried out to validate the correctness of rules in Table~\ref{tab:translation}. At the end of the section, we present an analysis of the obtained results. For the sake of brevity, we only report a partial translation of the workflow depicted in Figure~\ref{fig:workflow}, starting from the second conditional block marked with a  $?$. 
We introduce the following atomic propositions to represent activities: $\oplus$ for modeling the conditional block, $\rVert_{start}$ and $\rVert_{end}$ for activities defining starting and ending point of the join block and $\bill$ for \textit{Billing}. 
Similarly, we introduce a proposition for all other activities.
We use $t_1$ to indicate the transition reaching \textit{Billing} which starts from $\rVert_{start}$ and $t_2$ to indicate the transition reaching \textit{Shipping} which starts from $\rVert_{start}$. Finally, we use $sf$, $hf$ and $tf$ to model exceptions \textit{SoftwareFailure}, \textit{HardwareFailure} and \textit{TransportFailure}.
Formulae in the first row of Figure~\ref{fig:example} are the translation of the conditional block, the parallel block and \textit{Billing} activity.




\begin{figure*}
\begin{tabular}{c}
\begin{math}
\begin{array}{ccc}
\begin{gathered}
\oplus \Rightarrow (\oplus \land \neg (t_{yes}\lor t_{no})) \U (t_{yes}\lor t_{no})) \\
t_{yes} \Rightarrow \Y{\oplus} \land \neg \oplus\\
t_{no} \Rightarrow \Y{\oplus} \land \neg \oplus\\
t_{yes} \Rightarrow \neg t_{no}\\ 
\oplus \Rightarrow \neg\Y{\oplus} \land \neg\X{\oplus}
\end{gathered} & 
\begin{gathered}
\rVert_{start} \Rightarrow (\rVert_{start} \land \neg t_{yes}) \Snc \; t_{yes}\\
t_{yes} \Rightarrow \X{\rVert_{start}} \land \neg \rVert_{start}\\
\rVert_{start} \Rightarrow (\rVert_{start} \land \neg (t_1\lor t_2)) \U (t_2\lor t_2)) \\
t_2 \Rightarrow \Y{\rVert_{start}} \land \neg \rVert_{start}\\
t_1 \Rightarrow \Y{\rVert_{start}} \land \neg \rVert_{start}\\
t_1 \iFF t_2\\
\rVert_{start} \Rightarrow \neg\Y{\rVert_{start}} \land \neg\X{\rVert_{start}}
\end{gathered} &
\begin{gathered}
\bill \Rightarrow (\bill \land \neg t_1) \Snc \; t_1\\
t_1 \Rightarrow \X{\bill} \land \neg \bill\\
\bill \Rightarrow (\bill \land \neg t_1) \U \; t_3 \\
t_3 \Rightarrow \Y{\bill} \land \neg \bill
\end{gathered}
\\ \\
\begin{gathered}
(\bill \land \mathit{tf} \land \neg \mathit{Reject_2}) \Rightarrow \G{\bill}\\
(\mathit{hf} \Rightarrow 
(\neg \G{\mathit{hf}} \iFF \mathit{false}) \\
(\mathit{sf} \Rightarrow 
(\neg \G{\mathit{sf}} \iFF (\mathit{sf}) \U (\mathit{Recovery}))
\end{gathered}
&
\begin{gathered}
\G{\bill} \Rightarrow \F{
(\bill) \Snc(\bill\land (\mathit{tf} \land \neg \mathit{Reject_2})
}\\
\G{\bill} \Rightarrow  \F{
\begin{gathered}
\G{\bill \land \mathit{hf}} \lor \\
\G{\bill \land \mathit{sf} \land \neg\mathit{Recovery}} \lor \\
\G{\ship \land \mathit{hf}} 
\end{gathered} 
 }\\
\bill \land \mathit{sf} \land \neg\mathit{Recovery} \Rightarrow \G{\bill}
\end{gathered}&
\begin{gathered}
\mathit{tf} \Rightarrow \neg \X{\mathit{tf}} \\
\mathit{tf} \Rightarrow \ship \\
\bill \land \mathit{hf} \Rightarrow \G{\bill}
\end{gathered}
\end{array} 
\end{math}
\end{tabular}
\caption{some formulae modeling the portion of workflow in Figure~\ref{fig:workflow}. All formulae are conjuncted (symbol $\land$ is omitted for brevity) and globally quantified over time by $\mathbf{G}$. Observe that, since the workflow does not have an activity catching exception $\mathit{hf}$, whenever it occurs at the same time as activity $\bill$ then $\bill$ enters in an error state, i.e., infinite looping.}
\label{fig:example}
\end{figure*}

To validate the LTL model, we exploit Bounded Satisfiability Checking (BSC) \cite{PMS12} approach.
The key idea behind BSC is to build a finite representation, of length $k$, of an infinite ultimately periodic LTL model of the form $\alpha\beta^\omega$, where $\alpha$ and $\beta$ are finite words over the alphabet $2^{AP}$.
BSC tackles the complexity of checking the satisfiability for LTL formulae by avoiding the unfeasible construction of classical B\"uchi automata \cite{vw}.
\cite{BFRS11} proves that BSC problem for LTL and its extension is complete and that it can be reduced to a decidable Satisfiability Modulo Theory (SMT) problem.

All our tests were carried out by using \zot{} and the well and widely known Microsot Z3 SMT solver \cite{z3}.
\zot{} is a Bounded Model/Satisfiability checker, written in Lisp, that takes as input specifications written in a variety of temporal logics, and determines whether they are satisfiable or not. It performs the checks by encoding temporal logic formulae into the input language of various solvers, in particular SAT and SMT solvers. SAT solvers are capable of taking, as input, formulae written in propositional logic. SMT solvers instead accept formulae written in logics (fragments of First-Order Logic) that are richer than the simple propositional one. Suitable modules, interface \zot{} with the underlying SAT or SMT solvers. \zot{} scripts, which contain both the model to be analysed and the necessary commands to invoke the desired solver, are a collection of Lisp statements.

Table \ref{tb:tests} shows time -- in seconds -- required by \zot{} to verify a set of functional user-defined properties, memory occupation -- in MBytes -- and the result, i.e. whether the property is satisfied or not. 
Let $S$ be the formula which translates the model of our workflow in Figure~\ref{fig:workflow}. If $S$ is fed to \zot{} "as is", \zot{} will look for one of its execution; if it does not find one -- i.e., if the model is unsatisfiable -- then our translation is contradictory, hence it contains some flaws.
Now, let $P$ be one of the LTL formulae which model one of the functional properties we want to check on the workflow. If $S \land \neg P$ is unsatisfiable, this means that there is no execution that satisfies the workflow, that also satisfies $\neg P$, that is, that violates the property $P$, so $P$ holds; otherwise this means that there is at least one execution that satisfies both $S$ and $\neg P$; that is, there is at least one execution of the workflow that violates the property, so the property does not hold. If \zot{} determines that a formula is satisfiable, then the tool produces an execution that satisfies it that designer can use to check the correctness of the modeling, i.e. a counterexample \textit{trace} that is compatible with the workflow model but that violates the property.

With the introduction of exceptions, designers can simulate different scenarios, specifying and verifying functional properties to control the behavior of a workflow, when different kinds of exceptions occur. In the following, some examples of such type of LTL functional properties, which refer to the workflow in Figure~\ref{fig:workflow}, are defined to show the validity of the LTL based-semantics approach:

\begin{equation}\label{Property1}
\G{\neg tf \land \neg hf \land \neg sf} \Rightarrow \F{end}
\end{equation}

Formula \eqref{Property1} states that, if no exceptions occur, the workflow must terminate. Feeding the model checker with the negation of this property and the LTL model of the workflow, we can check if \textbf{all} possible executions of the workflow reach \textit{end} state. As reported in Table~\ref{tb:tests}, the property holds, since there are no traces which satisfy its negation. 

\begin{equation}\label{Property2}
(\G{\neg tf \land \neg sf} \land \F{hf})  \Rightarrow \F{end}
\end{equation}

Formula \eqref{Property2} is needed to verify if the workflow terminates or not, when an \textit{HardwareFailure} exception occurs. \textit{HardwareFailure} exception is thrown during execution of \textit{InternalCreditCheck} activity, which models an hardware fault of the machine that executes the software needed to automatically check the credit of the customer. As reported in Table~\ref{tb:tests}, property \eqref{Property2} does not hold: in this case \zot{} returns an execution trace which is a counterexample. By analyzing the trace -- which is not shown for brevity -- we can note that activities \textit{Billing} and \textit{Shipping} loop forever, since $hf \in probe(Billing)$ and also $hf \in probe(Shipping)$. In particular, the trace shows that at least a path exists where there are no activities which catch $hf$ exception between \textit{InternalCreditCheck} and \textit{Billing} or \textit{Shipping}, which means that error exceptions are not properly handled. As we remark in section \ref{section-casestudy}, it is possible to refine the workflow, introducing a state similar to \textit{SoftwareRecovery} to catch $hf$ exception.   

%


\begin{equation}\label{Property4}
\F{\G{\bill}} \Leftrightarrow \F{\G{Ship}}
\end{equation}

\begin{equation}\label{Property5}
\F{\G{\bill}} \Leftrightarrow \F{\G{Arch}}
\end{equation}

Formulae \eqref{Property4} and \eqref{Property5} model different type of properties respect to the other ones: Formula \eqref{Property4} states that activity \textit{Billing} loops forever if and only if activity \textit{Shipping} do the same,  while Formula \eqref{Property5} states the same for the activities \textit{Billing} and \textit{Archiving}. \textit{Ship} and \textit{Arch} are the abbreviations respectively of \textit{Shipping} and \textit{Archiving}. Table~\ref{tb:tests} shows that the property modeled by Formula \eqref{Property4} holds, while the other one does not hold: in fact, analyzing the counterexample trace and the workflow, we can note that if activity \textit{Billing} loops forever, it is not possible to reach activity \textit{Archiving}. 


 
All tests have been carried out on a 3.3 Ghz quad core PC with 16 Gbytes of Ram. 
The bound $k$, which is a user-defined parameter that corresponds to the maximal length of runs analysed by \zot{}, corresponds to the number of steps needed to build the bounded representation of the model. The value chosen is $k=35$.

\setlength{\tabcolsep}{4pt}
\begin{table}[!hbt]
\begin{center}
\caption{Test results.}
\label{tb:tests}
\begin{tabular}{cccc}
Formula                            & Time (s) & Memory (Mb) & Result \\\hline
Formula \eqref{Property1} & 2.836  &   16    &  UNSAT \\
Formula \eqref{Property2} & 3.883  &   60    &  SAT \\
Formula \eqref{Property4} & 3.843  &   60    &  UNSAT \\
Formula \eqref{Property5} & 3.785  &   60    &  SAT \\
\end{tabular}
\end{center}
\end{table}

As it is expected, although the model is composed of a number of atomic propositions which is equal to the sum of the number of states and transitions of the workflow (about 50) plus the atomic propositions needed to model exceptions, the quantity of time and the amount of memory needed to perform the analysis are very low. In particular, to verify properties like the ones modeled by Formulae \eqref{Property1} and \eqref{Property4}, \zot{} must exhaustively analyse all possible runs to return UNSAT, which is the worst case in terms of time and memory consumed; taking it into account, we can conclude that it is feasible, using modern model checking tools such as \zot{}, to perform formal verification of non-trivial functional properties, in a limited amount of resources, allowing designer to execute the analysis in an interactive real-time manner.

We left as future work the implementation and testing of more interesting properties, such as hard real-time properties. 

\section{Conclusions and future work}
\label{section-conclusions}

The major objective of this paper is demonstrating how temporal logics are effective in giving semantics and iteratively enforce requirements into the process. Our approach is lightweight and allows the reuse of existing tool support. The working assumption is that a lightweight solution would easily fit into processes that are already in place without the need for a radical change of procedures, tools and people's attitudes. The complexity of formalisms and invasiveness of methods have been indeed demonstrated to be one of the major drawback and obstacle for deployment of formal engineering techniques into mundane projects. The case study is a purchase workflow, but the results can be extended to other systems with emphasis on dependability and abnormal behavior management. The treatment of exception handling, and more in general of recovery, is another substantial contribution that has been less frequently investigated with similar techniques and tools.

The workflow patterns here analyzed are limited with respect to a real scenario. Workflow patterns as presented in \cite{wmp2003workflow} need to be investigated and encoded. Once workflows are intended as graphs and transitions are treated like in this paper, similarities emerge with the Petri Nets approach, in particular with Workflow Petri Nets \cite{Aalst:1997}. 

Future work aims at extending the current translation of workflows by using more expressive logics. In particular, we plan to extend the basic definition of workflow by adding timing constraints on activities and transitions. To model timed workflow we may exploit CLTLoc \cite{BRS13b}, which is an LTL based logic where atomic formulae are both atomic propositions and constraints over dense clocks. Zero-time modelization is also an open issue. When some workflow activities have a negligible duration with respect to the other ones, they may be modeled as having a logical zero time duration. This implies Zeno behaviours and other counterintuitive consequences. \cite{FMMR12} introduces a new metric temporal logic called \textit{X-TRIO}, which exploits the concepts of \textit{Non-Standard Analysis} \cite{R96}. The way to "glue" together CLTLoc with X-TRIO is a promising research strand.

Finally, runtime evolution in business processes \cite{Panzica14} and, more in general, the idea of self-reconfiguring systems are related issues we intend to further explore.


\subsubsection*{Acknowledgements}
The authors acknowledge the support and advice given by Anirban Bhattacharyya, Alexandr Naumchev, Diego P\'erez and all the other Friends at Politecnico di Milano.



\bibliographystyle{apalike}
{\small
\bibliography{bibliography}}

\end{document}